\documentclass[reprint,prd,twocolumn,superscriptaddress,showpacs,showkeys,floatfix,preprintnumbers]{revtex4-2}

\pdfoutput=1

\usepackage[english]{babel}
\usepackage[utf8]{inputenc}
\usepackage[T1]{fontenc}
\usepackage{amsmath,amssymb,braket,slashed,mathrsfs}
\usepackage{pifont}
\usepackage{hyperref}
\usepackage{color,hyperref}
\usepackage{cleveref}
\usepackage{amsmath}
\usepackage{graphicx}
\usepackage{booktabs}
\usepackage{multirow}
\usepackage{subfigure}
\usepackage{float}

\newcommand{\beq}{\begin{eqnarray}}
\newcommand{\eeq}{\end{eqnarray}}
\newcommand{\be}{\begin{equation}}
\newcommand{\ee}{\end{equation}}
\newcommand{\ba}{\begin{eqnarray}}
\newcommand{\ea}{\end{eqnarray}}
\newcommand{\nn}{\nonumber}
\newcommand{\bit}{\begin{itemize}}
\newcommand{\eit}{\end{itemize}}
\newcommand{\rw}{\rightarrow}

\newcommand{\zzuphy}{School of Physics and Laboratory of Zhongyuan Light, Zhengzhou University, Zhengzhou, Henan 450001, China}
\newcommand{\innovation}{Collaborative Innovation Center of Quantum Matter, Beijing 100871, China}
\newcommand{\chep}{Center for High Energy Physics, Peking University, Beijing 100871, China}
\newcommand{\pkuphy}{School of Physics, Peking University, Beijing 100871,
China}

\newcommand{\IHEP}{Institute of High Energy Physics, Chinese Academy of Sciences, Beijing 100049, People’s Republic of China}

\begin{document}

\title{Lattice QCD calculation of $D_s^{*}$ radiative decay with (2+1)-flavor Wilson-clover ensembles}

\author{Yu Meng}
\email[Email: ]{yu_meng@zzu.edu.cn}
\affiliation{\zzuphy}
\author{Jin-Long Dang}
\affiliation{\pkuphy}
\author{Chuan Liu} 
\email[Email: ]{liuchuan@pku.edu.cn}

\affiliation{\pkuphy}\affiliation{\chep}\affiliation{\innovation}
\author{Zhaofeng Liu}
\affiliation{\IHEP}\affiliation{\chep}
\author{Tinghong Shen}
\affiliation{Hubei Nuclear Solid Physics Key Laboratory, School of Physics and Technology, Wuhan University, Wuhan, Hubei 430072, People’s Republic China}\affiliation{\IHEP}
\author{Haobo Yan}
\affiliation{\pkuphy}
\author{Ke-Long Zhang}
\affiliation{Computer Network Information Center, Chinese Academy of Sciences}

\date{\today}

\begin{abstract}
We perform a lattice calculation on the radiative decay of $D_s^*$ using the (2+1)-flavor Wilson-clover gauge ensembles generated by CLQCD collaboration. A method allowing us to calculate the form factor with zero transfer momentum is proposed and applied to the radiative transition $D_s^*\rightarrow D_s\gamma$ and the Dalitz decay $D_s^*\rightarrow D_s e^+e^-$. After a continuum extrapolation using three lattice spacings, we obtain $\Gamma(D_s^*\rightarrow D_s \gamma)=0.0549(54)$ keV, where the error is purely statistical. The result is consistent with previous lattice calculations but with a error reduced to only a fifth of the before. The Dalitz decay rate is also calculated for the first time and the ratio with the radiative transition is found to be $R_{ee}=0.624(3)\%$. A total decay width of $D_s^*$ can then be determined as 0.0587(54) keV taking into account the experimental branching fraction. Combining with the most recent experimental measurement on the branching fraction of the purely leptonic decay $D_s^{+,*}\rightarrow e^+\nu_e$, we obtain the quantity $f_{D_s^*}|V_{cs}|=(190.5^{+55.1}_{-41.7_{\textrm{stat.}}}\pm 12.6_{\textrm{syst.}})$ MeV, where the stat. is only the statistical error from the experiment, and syst. results from the experimental systematic uncertainty and the lattice statistical error. Our result leads to an improved systematic uncertainty compared to $42.7_{\textrm{syst.}}$ obtained using previous lattice prediction of total decay width $0.070(28)$ keV as the input.

\end{abstract}

\maketitle

\section{Introduction}

Testing the standard model precisely and searching for signals or even hints for new physics beyond the standard model is one of the major goals of contemporary particle physics. Various flavor-changing weak decay processes can be used to extract relevant Cabibbo-Kobayashi-Maskawa(CKM) matrix elements and then test them under 3 flavor unitarity, which has become a well-known and extremely important  direction in flavor physics~\cite{FlavourLatticeAveragingGroupFLAG:2021npn}. 
For the electromagnetic decays without involving the CKM matrix elements, the decay rate can be measured experimentally and also be calculated theoretically, hence providing an alternative and even more direct way to test the standard model. In recent works on $\eta_c\rightarrow 2\gamma$~\cite{Meng:2021ecs,Colquhoun:2023zbc}, for example, the decay rate has been calculated precisely and it appears to differ significantly from the Particle Data Group’s reported value. It therefore leaves some interesting physics in this channel.

In this work, we focus on the radiative decay of the excited strange charm meson, the vector meson $D_s^*$ with quark content $c\bar{s}$. Though the particle mass and the branching fraction of the known decay channels have been measured, the total decay width of $D_s^*$ is not determined experimentally. A possible way to extract the total decay width is to combine the calculated partial decay width, 
such as $D_s^*\rightarrow D_s\gamma$ and its experimental branching fraction. On the theoretical side, an impressive achievement comes from the lattice calculation~\cite{Donald:2013sra}, in which the authors obtained the radiative decay width as $\Gamma(D_s^*\rightarrow D_s \gamma)=0.066(26)$ keV. It thereby determines the total decay width $\Gamma_{D_s^*}^{\textrm{total}}=0.070(28)$ keV. A most recent experimental measurement on the branching fraction of the purely leptonic decay $D_s^{*,+}\rightarrow e^+\nu_e$ gives $\operatorname{Br}(D_s^{*,+}\rightarrow e^+\nu_e)=(2.1^{+1.2}_{-0.9_{\textrm{stat.}}}\pm 0.2_{\textrm{syst.}})\times 10^{-5}$~\cite{BESIII:2023zjq}. Combining this branching fraction and the total decay width, it obtains $f_{D_s^*}|V_{cs}|=(207.9^{+59.4}_{-44.6_{\textrm{stat.}}}\pm 42.7_{\textrm{syst.}})$ MeV. The systematic uncertainty comes from the uncertainties in the experimental measurement and the total decay width. Since the precision of the theoretical total decay width is about 40\%, significantly larger than the experimental 10\%, it is therefore urgent to reduce the theoretical uncertainty for a more precise extraction of the quantity $f_{D_s^*}|V_{cs}|$.

The aim of this work is to further
improve upon the previous lattice study of this radiative decay. Several improvements are made to obtain a more accurate
result. i)  We adopt a novel method to extract the on-shell transition factor. The new method allows a calculation of an off-shell transition factor with zero transfer momentum. When the continuous momentum extrapolation is performed, the accuracy of the on-shell factor is well controlled by this point; ii) We consider a large number of time separations between the initial and final particles in our calculation. A correlated fit to a constant at large time separation is performed and the excited-state contamination is well removed; iii) We utilize three gauge ensembles with different lattice spacings, the finest of which is 0.052 fm, leading to a well-controlled continuous limit $a^2\rightarrow 0$. These efforts finally enable us to obtain the decay width with a statistical precision of about 9.8\%.

The rest of this paper is organized as follows. In Sec.~\ref{sec:method}, we
introduce the methodology utilized in this work for calculating the radiative decay width. This section is divided into three parts: in Sec.~\ref{sec:scalar_method} the theoretical framework is given; in Sec.~\ref{sec:relation_M_E} the hadronic function is extracted from the lattice data; in Sec.~\ref{sec:decay_width} the decay widths of $D_s^*\rightarrow D_s^*\gamma$ and $D_s^*\rightarrow D_s e^+e^-$ are obtained using the form factors calculated on the lattice. In Sec.~\ref{sec:result} we give details of the simulations and show the main results. This section is further divided into three parts: in Sec.~\ref{sec:mass_spectrun} the numerical values of $D_s^*$ and $D_s$ masses, together with the dispersion relation of $D_s$ particle are presented; in Sec.~\ref{sec:radiative} the results of $D_s^*\rightarrow D_s \gamma$ are obtained, a continuum extrapolation under three lattice spacings is performed; 
in Sec.~\ref{sec:dalitz} the results of $D_s^*\rightarrow D_s e^+e^-$ are summarized; Finally, we conclude in Sec.~\ref{sec:conclude}.

\section{Methodology}
\label{sec:method}
The lattice study on the radiative transition process is quite mature, either for the traditional momentum extrapolation, or the latter twisted boundary condition~\cite{Bedaque:2004kc,deDivitiis:2004kq}. We will not go into these details in this paper. Instead, we would proceed in another way, which is called the scalar function method in recent years. Such a method has been widely applied to various physical processes~\cite{Feng:2019geu,Feng:2020zdc,Tuo:2021ewr,Meng:2021ecs,Zou:2021mgf,Fu:2022fgh,Tuo:2022hft,Christ:2022rho,Meng:2023bjc} and achieved great successes. For more detailed derivations in this paper, we refer to the supplementary materials in our previous study on the charmonium two-photon decay\cite{Meng:2021ecs}, where the same parameterization of the form factor is utilized.

\subsection{Scalar function method}\label{sec:scalar_method}
We start with a Euclidean hadronic function in the infinite volume
\be
\label{eq:H-def}
H_{\mu\nu}(\vec{x},t)= \langle 0|\mathcal{O}_{D_s}(\vec{x},t)J_{\nu}^{\textrm{em}}(0)|D_{s,\mu}^*(p')\rangle, \quad t>0\;,
\ee
where $J_\nu^{em}=\sum_qe_q\,\bar{q}\gamma_\nu q$ ($e_q=2/3,-1/3,-1/3,2/3$ for $q=u,d,s,c$. The $|D_{s,\mu}^*(p')\rangle$ is a $D_{s}^*$ state with momentum $p'=(im_{D^*_s},\vec{0})$ and $\mathcal{O}_{D_s}$ is the interpolating operator of $D_s$. At large time $t$, the hadronic function is saturated by the single $D_s$ state
\beq
&&H_{\mu\nu}(x)\doteq H_{\mu\nu}^{D_s}(x)=\int \frac{d^3\vec{p}}{(2\pi)^3}\frac{1}{2E_{D_s}}e^{-E_{D_s}t+i\vec{p}\cdot \vec{x}} \nonumber \\
&&\times \langle 0|\mathcal{O}_{D_s}(0)|D_s(\vec{p})\rangle \langle D_s(\vec{p})|J_{\nu}(0)|D_{s,\mu}^*(p')\rangle
\eeq
Considering the following parametrizations
\beq\label{eq:F_param}
\langle 0|\mathcal{O}_{D_s}(0)|D_s(\vec{p})\rangle &=&Z_{D_s} \nonumber \\
\langle D_s(p)|J_{\nu}^{\textrm{em}}(0)|D_{s,\mu}^*(p')\rangle&=&\frac{2V_{\textrm{eff}}(q^2)}{m_{D_s}+m_{D_s^*}}\epsilon_{\mu\nu\alpha \beta}p_{\alpha}p'_{\beta}
\eeq
where an effective transition factor $V_{\textrm{eff}}(q^2)$ is introduced and the square of transfer momentum $q^2$ is determined by $q^2=(m_{D_s^*}-E_{D_s})^2-|\vec{p}|^2$ as $D_s^{*}$ is at rest. Then, the spatial Fourier transform of $H_{\mu\nu}(\vec{x},t)$ yields
\beq\label{eq:V_traditional}
\tilde{H}_{\mu\nu}(\vec{p},t)&&\doteq \tilde{H}_{\mu\nu}^{D_s}(\vec{p},t)=\frac{1}{m_{D_s}+m_{D_s^*}}\frac{Z_{D_s}}{E_{D_s}}e^{-E_{D_s}t} \nonumber \\
&&\times \epsilon_{\mu\nu\alpha \beta}p_{\alpha}p'_{\beta}V_{\textrm{eff}}(q^2)
\eeq

The conventional way to extract the form factor $V_{\textrm{eff}}(q^2)$ is to utilize the above equation at a series of nonzero lattice momentum $\vec{p}=2 \pi \vec{n}/L,\vec{n}\neq 0$. The on-shell transition factor $V_{\textrm{eff}}(0)$ is obtained by a momentum extrapolation with these data at discrete $q^2$ as inputs. We remark that the specific momentum point $\vec{p}=0$ closest to the on-shell condition $q^2=0$ is missed in such a way. It is easy to check that $\tilde{H}_{\mu\nu}(\vec{p},t) = 0$ if the $D_s$ and $D_s^*$ are both stationary, and hence the $\vec{p}=0$ data cannot be incorporated into the analysis via this quantity. Therefore, an extrapolation including this point will improve the precision and, more importantly, the reliability of the momentum extrapolation. To achieve this, we 
construct a scalar function given below
\beq
\mathcal{I}(t,|\vec{p}|)&=&\frac{1}{m_{D_s^*}|\vec{p}|^2}\epsilon_{\mu\nu\alpha \beta}p_{\alpha}p'_{\beta}\tilde{H}_{\mu\nu}(\vec{p},t)
\eeq
This quantity, on the one hand, is related to the transition factor $V_{\textrm{eff}}(q^2)$ via
\be
\mathcal{I}(t,|\vec{p}|)=\frac{-2Z_{D_s}m_{D_s^*}}{m_{D_s}+m_{D_s^*}}V_{\textrm{eff}}(q^2)\frac{e^{-E_{D_s}t}}{E_{D_s}}\;;
\ee
On the other hand, it can be computed directly only using the hadronic function $H_{\mu\nu}(\vec{x},t)$ defined in Eq.~(\ref{eq:H-def}) as an input,
\beq
\mathcal{I}(t,|\vec{p}|)=\int d^3\vec{x}\frac{j_1(|\vec{p}||\vec{x}|)}{|\vec{p}||\vec{x}|}\epsilon_{\mu\nu \alpha 0}x_{\alpha}H_{\mu\nu}(\vec{x},t)
\eeq
where $j_1(x)$ is the spherical Bessel function. 
This finally leads to,
\beq\label{eq:V_eff}
V_{\textrm{eff}}(q^2)&=&\frac{-(m_{D_s}+m_{D_s^*})E_{D_s}}{2Z_{D_s}m_{D_s^*}} e^{E_{D_s}t} \nonumber \\
&\times& \int d^3\vec{x}\frac{j_1(|\vec{p}||\vec{x}|)}{|\vec{p}||\vec{x}|}\epsilon_{\mu\nu \alpha 0}x_{\alpha}H_{\mu\nu}(\vec{x},t)\;.
\eeq

It is easy to verify the momentum $|\vec{p}|=0$ is immediately accessible since $j_1(x)/x $ tends to a finite value as $x \rightarrow 0$. The on-shell
transition factor $V_{\textrm{eff}}(0)$ can be determined by a general polynomial extrapolation,
\be\label{eq:mom_extra}
V_{\textrm{eff}}(q^2)=d_0+d_1 \cdot \frac{q^2}{m_{D_s^*}^2}+d_2\cdot \frac{q^4}{m_{D_s^*}^4}+\mathcal{O}(q^6/m_{D_s^*}^6)
\ee
where the coefficients $d_i$ are introduced and $V_{\textrm{eff}}(0)\equiv d_0$. Since $q^2=(m_{D_s^*}-m_{D_s})^2\equiv (\delta m)^2$ as $\vec{p}=0$, the difference of $V_{\textrm{eff}}(0)$ and $V_{\textrm{eff}}((\delta m)^2)$ is thereby a very small quantity with the consideration of $(\delta m)^2/m_{D_s^*}^2 \sim 0.46\%$. It is therefore expected that the extrapolation precision with $\vec{p}=0$ included can be significantly improved.

\subsection{Hadronic function $H_{\mu\nu}(\vec{x},t)$}\label{sec:relation_M_E}

The hadronic function $H_{\mu\nu}(\vec{x},t)$ can be
extracted from a three-point function $C_{\mu\nu}^{(3)}(\vec{x},t)$
\beq
C_{\mu\nu}^{(3)}(\vec{x},t)=\langle \mathcal{O}_{D_s^+}(\vec{x},t)J_{\nu}^{\textrm{em}}(0)\mathcal{O}_{D_{s,\mu}^{+,*}}^{\dagger}(-t) \rangle
\eeq
where interpolating operators are chosen as $\mathcal{O}_{D_{s,\mu}^{+,*}}^{\dagger}=-\bar{c}\gamma_{\mu}s$ and $\mathcal{O}_{D_s^+}=\bar{s}\gamma_5 c$. In this work, we only consider the connected contributions. Then, it has the following quark contractions:
\beq\label{eq:3pt_lat}
&&C_{\mu\nu}^{(3)}(\vec{x},t)= \nonumber \\
&&-e_c\langle \gamma_5\gamma_{\mu}S_{s}^{\dagger}(\vec{x},t;-t)S_{c}(\vec{x},t;0)\gamma_{\nu}S_{c}(0,-t) \rangle \nn \\
&&+e_s \langle \gamma_5\gamma_{\mu}S_s^{\dagger}(0,-t)\gamma_{\nu}S_{s}^{\dagger}(\vec{x},t;0)S_c(\vec{x},t;-t)\rangle
\eeq
Thus, the hadronic function $H_{\mu\nu}(x)$ is given by
\beq
H_{\mu\nu}(\vec{x},t)=\frac{2m_{D_s}}{Z_{D_s}}e^{m_{D_s}t}C_{\mu\nu}^{(3)}(\vec{x},t)
\eeq
where $m_{D_s}$ and $Z_{D_s}$ are extracted from the two-point function $C^{(2)}(\vec{p},t)=\sum\limits_{\vec{x}}\cos(\vec{p}\cdot \vec{x})\langle\mathcal{O}_{h}(\vec{x},t)\mathcal{O}_{h}^{\dagger}(0)\rangle $ by a single-state fit
\beq\label{eq:2pt}
C^{(2)}(\vec{p},t)=\frac{Z_{h}^2}{2E_{h}}\left( e^{-E_{h} t}+e^{-E_{h}(T-t)} \right)
\eeq
with $m_{h}=E_{h}(\vec{p}=0)$ the ground-state energy and $Z_h=\langle h|\mathcal{O}^{\dagger}_h|0 \rangle$ is the overlap amplitude for the ground state. The symbol $h$ denotes the hadron, for example, $D_s$ or $D_s^{*}$ in this paper. For the computation of the three-point function $C_{\mu\nu}^{(3)}(\vec{x},t)$, we place the point source propagator on the current and wall source propagator on the initial hadron. All the propagators are produced on a large number of time slices by average to increase the statistics based on time translation invariance.

\subsection{Decay width of $D_s^*\rightarrow D_s\gamma$ and $D_s^*\rightarrow D_se^{+}e^{-}$}\label{sec:decay_width}
The amplitude of $D_s^{+,*}\rw \gamma D_s^+$ can be written as
\be\label{eq:amplitude}
i\mathcal{M}(\lambda',\lambda)=ie\epsilon^{\mu}_{D_s^{*}}(p',\lambda')\epsilon^{\nu}_{\gamma}(q,\lambda)\langle D_s(p)|J_{\nu}^{\textrm{em}}(0)|D_{s,\mu}^*(p')\rangle
\ee
where $\epsilon^{\mu}_{D_s^{*}}(p',\lambda')$ is the polarization vector of the $D_s^*$ particle and $\epsilon^{\nu}_{\gamma}(q,\lambda)$ is the photon polarization with the four-vector momentum $q=p'-p$. These polarizations satisfy the following identities:
\beq
\sum\limits_{\lambda}\epsilon^{\mu}_{D_s^*}(p',\lambda)\epsilon^{\nu}_{D_s^*}(p',\lambda)&=&-g_{\mu\nu}+\frac{p'_{\mu}p'_{\nu}}{m_{D_s^*}^2} \\
\sum\limits_{\lambda}\epsilon^{\mu}_{\gamma}(p,\lambda)\epsilon^{\nu}_{\gamma}(p,\lambda)&=&-g_{\mu\nu}
\eeq
Combining the parametrization in Eq.~(\ref{eq:F_param}), it immediately leads to the following decay width
\beq
&&\Gamma(D_s^{*}\rw \gamma D_s)\nonumber \\
&&=\frac{1}{2m_{D_s^*}}\int \frac{d^3\vec{q}}{(2\pi)^32|\vec{q}|}\int \frac{d^3\vec{p}}{(2\pi)^32E_{D_s}}\nonumber \\
&&\times (2\pi)^4\delta^{(4)}(p'-p-q) \times \frac{1}{3}\sum\limits_{\lambda'}\sum\limits_{\lambda}|\mathcal{M}(\lambda',\lambda)|^2 \nonumber \\
&&=\frac{4}{3}\frac{\alpha (\delta m)^3}{(m_{D_s}+m_{D_s^*})^2}|V_{\textrm{eff}}(0)|^2
\eeq
where $\alpha\equiv e^2/4\pi$. The factor $1/3$ in the third line denotes the average over 
three polarizations of $D^*_{s}$ in its rest frame and the final photon polarization 
has been summed up.

The amplitude of the Dalitz decay $D_s^*\rightarrow D_se^+e^-$ has additionally photon propagator $-ig_{\nu\nu'}/q^2$ and leptonic current $\bar{u}_e\gamma_{\nu'}u_e$ compared with the amplitude of the radiative decay $D_s^*\rightarrow D_s\gamma$ in Eq.~(\ref{eq:amplitude}). A direct calculation similar to the above gives the Dalitz decay widths normalized to the corresponding radiative decay as follows~\cite{LANDSBERG1985301}
\beq\label{eq:Ree}
R_{ee}&=&\frac{\Gamma(D_s^*\rw D_se^+e^-)}{\Gamma(D_s^*\rw D_s\gamma)} \nonumber \\
&=&\frac{\alpha}{3\pi}\int \frac{dq^2}{q^2}\Big{|}\frac{V_{\textrm{eff}}(q^2)}{V_{\textrm{eff}}(0)}\Big{|}^2\left(1-\frac{4m_e^2}{q^2}\right)^{\frac{1}{2}}\left(1+\frac{2m_e^2}{q^2}\right) \nonumber \\
&\times& \left[\left(1+\frac{q^2}{m_{D_s^*}^2-m_{D_s}^2}\right)^2-\frac{4m_{D_s^*}^2q^2}{(m_{D_s^*}^2-m_{D_s}^2)^2}\right]^{\frac{3}{2}}
\eeq

\section{Simulations and results}\label{sec:result}
\begin{table}[!h]
\begin{ruledtabular}
\begin{tabular}{cccc}
\textrm{Ensemble} & C24P29 & C32P30 & C48P32 \\
\hline
$a(\textrm{fm})$ & 0.10530(18) &0.07746(18) & 0.05187(26) \\
$a\mu_s$ & -0.2400 & -0.2050 &-0.1700 \\
$a\mu_c$ &0.4479 &0.2079 & 0.0581 \\
$L^3\times T$ & $24^3\times 72$ & $32^3\times 96$ & $48^3\times 144$ \\
$N_{\textrm{cfg}}\times N_{\textrm{src}}$ & $450\times 72$ & $377\times 96$ & $306\times 72$ \\
$m_{\pi}(\textrm{MeV})$ & 292.7(1.2) & 303.2(1.3) & 317.2(0.9) \\
$m_{J/\psi}(\textrm{MeV})$ & 3098.6(0.3) & 3094.9(0.4) & 3096.5(0.3) \\
$t$ & 3-18 & 2-22& 8-30 \\
$Z_V$ & 0.79814(23)& 0.83548(12)& 0.86855(04)
\end{tabular}
\end{ruledtabular}
\caption{
Parameters of gauge ensembles used in this work. From top to bottom, we list the ensemble name, the lattice spacing $a$,
the bare quark mass including the strange quark $a\mu_s$ and valence charm quark $a{\mu}_c$, the spatial and temporal lattice size $L$ and $T$,
the number of the measurements of the correlation function for each ensemble $N_{\textrm{cfg}}\times N_{\textrm{src}}$, the pion mass $m_{\pi}$, the $J/\psi$ mass $m_{J/\psi}$, the range of the time separation $t$ between the initial hadron and the electromagnetic current, and the vector normalization constant $Z_V$.
Here, $L$, $T$ and $t$ are given in lattice units.\label{table:cfg}}
\end{table}

\begin{figure*}[htbp]
\centering
\subfigure{
\centering
\includegraphics[width=5.5cm]{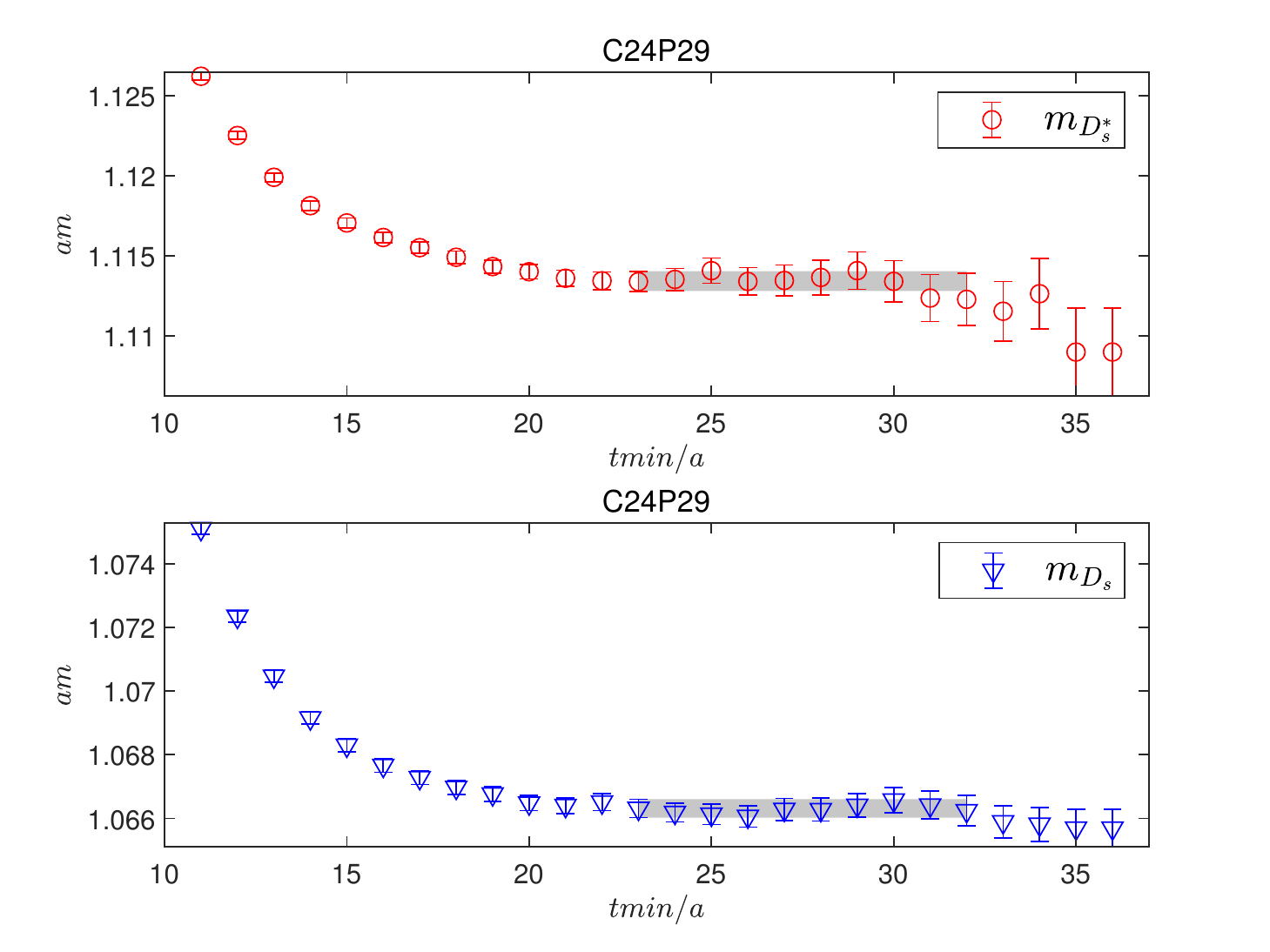}
}
\subfigure{
\centering
\includegraphics[width=5.5cm]{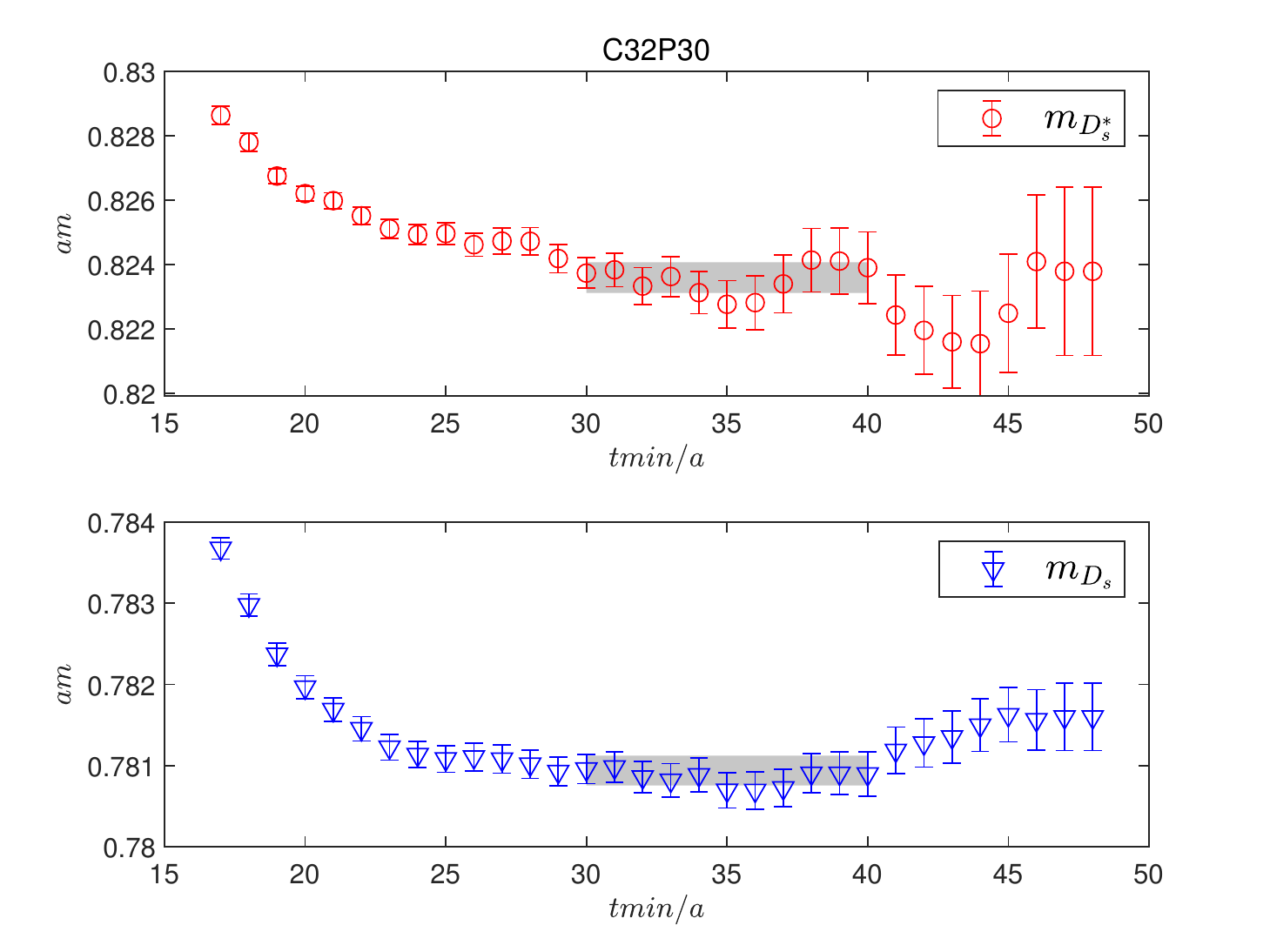}
}
\subfigure{
\centering
\includegraphics[width=5.5cm]{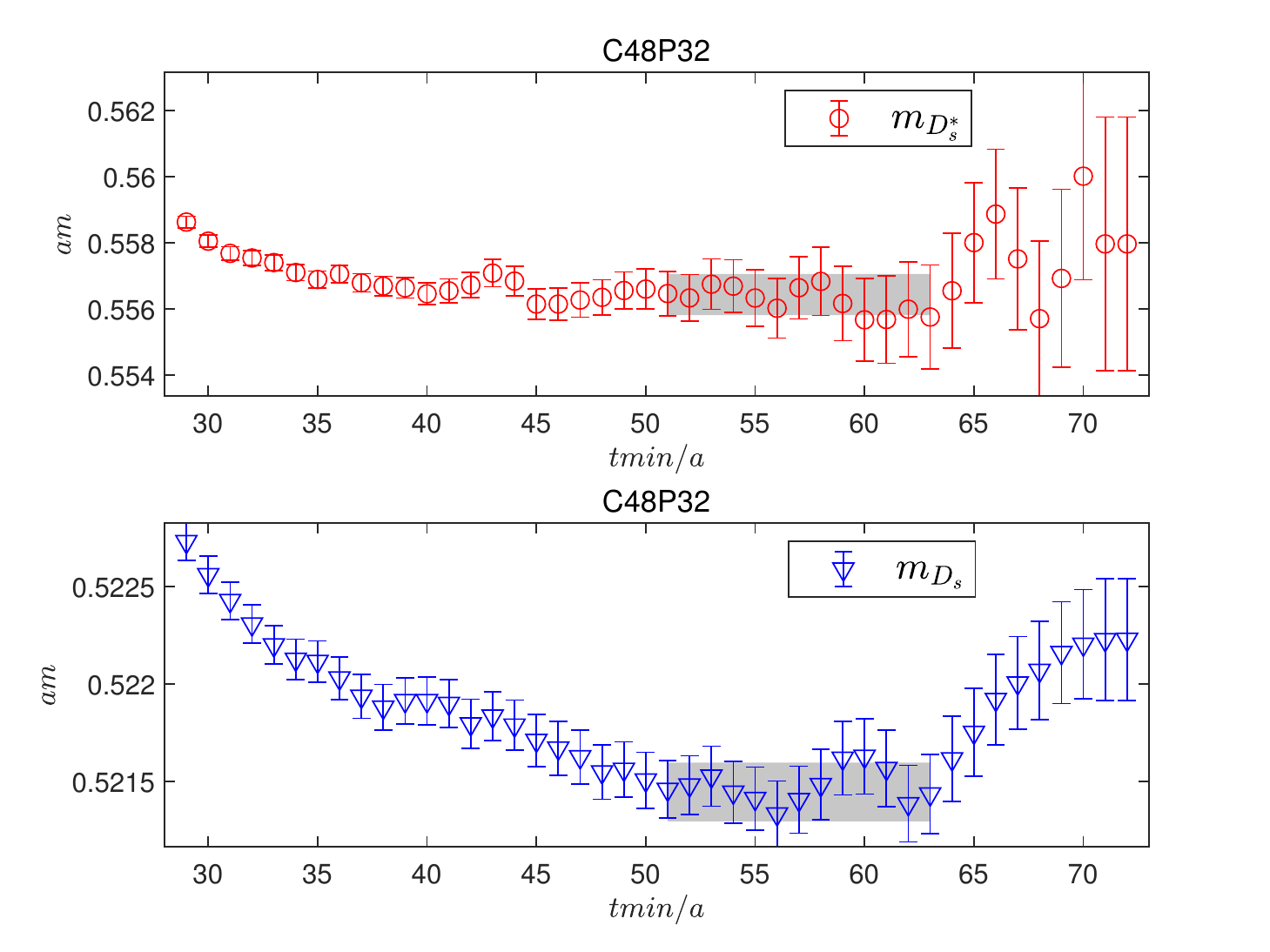}
}
\subfigure{
\centering
\includegraphics[width=5.5cm]{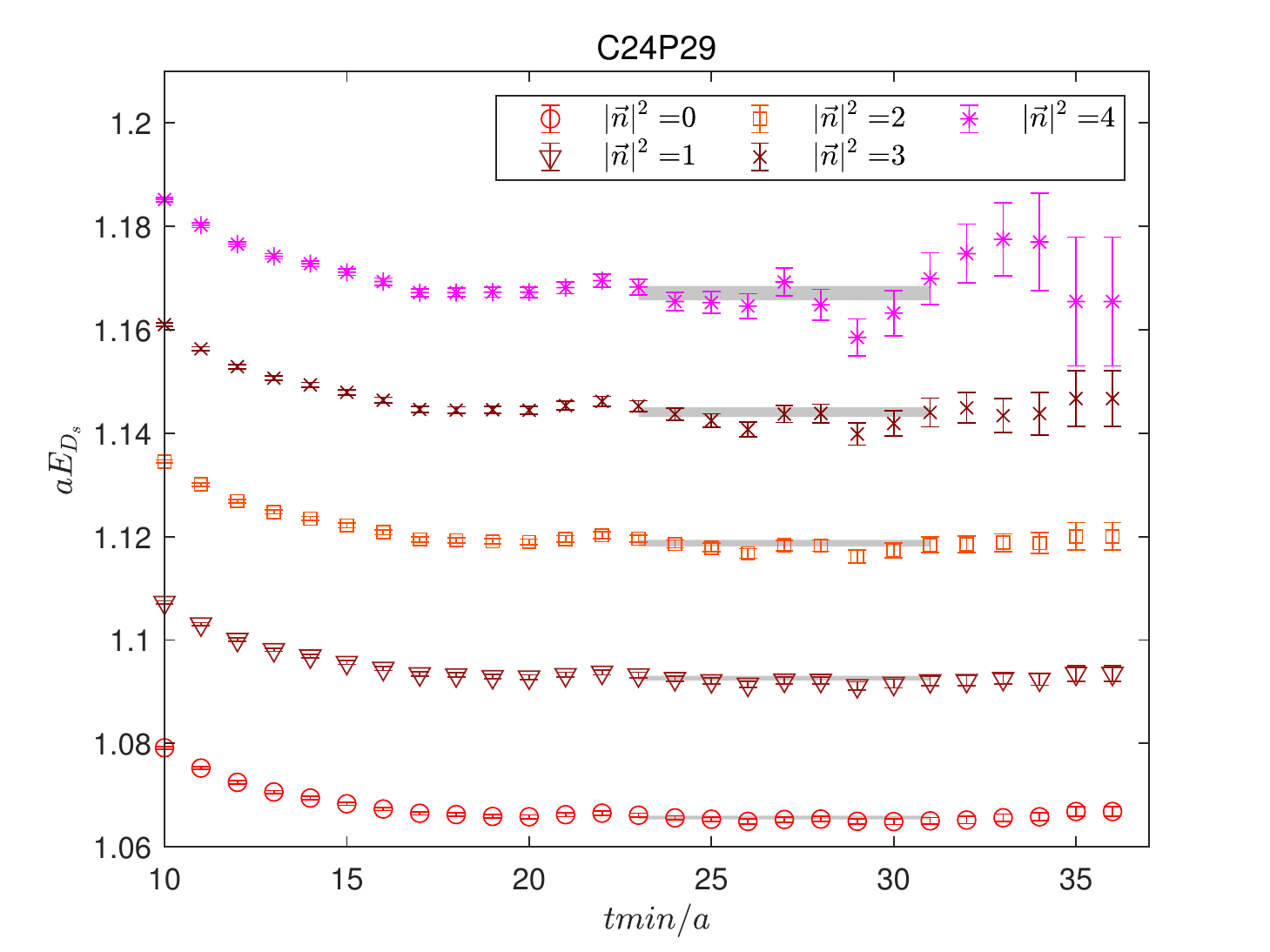}
}
\subfigure{
\centering
\includegraphics[width=5.5cm]{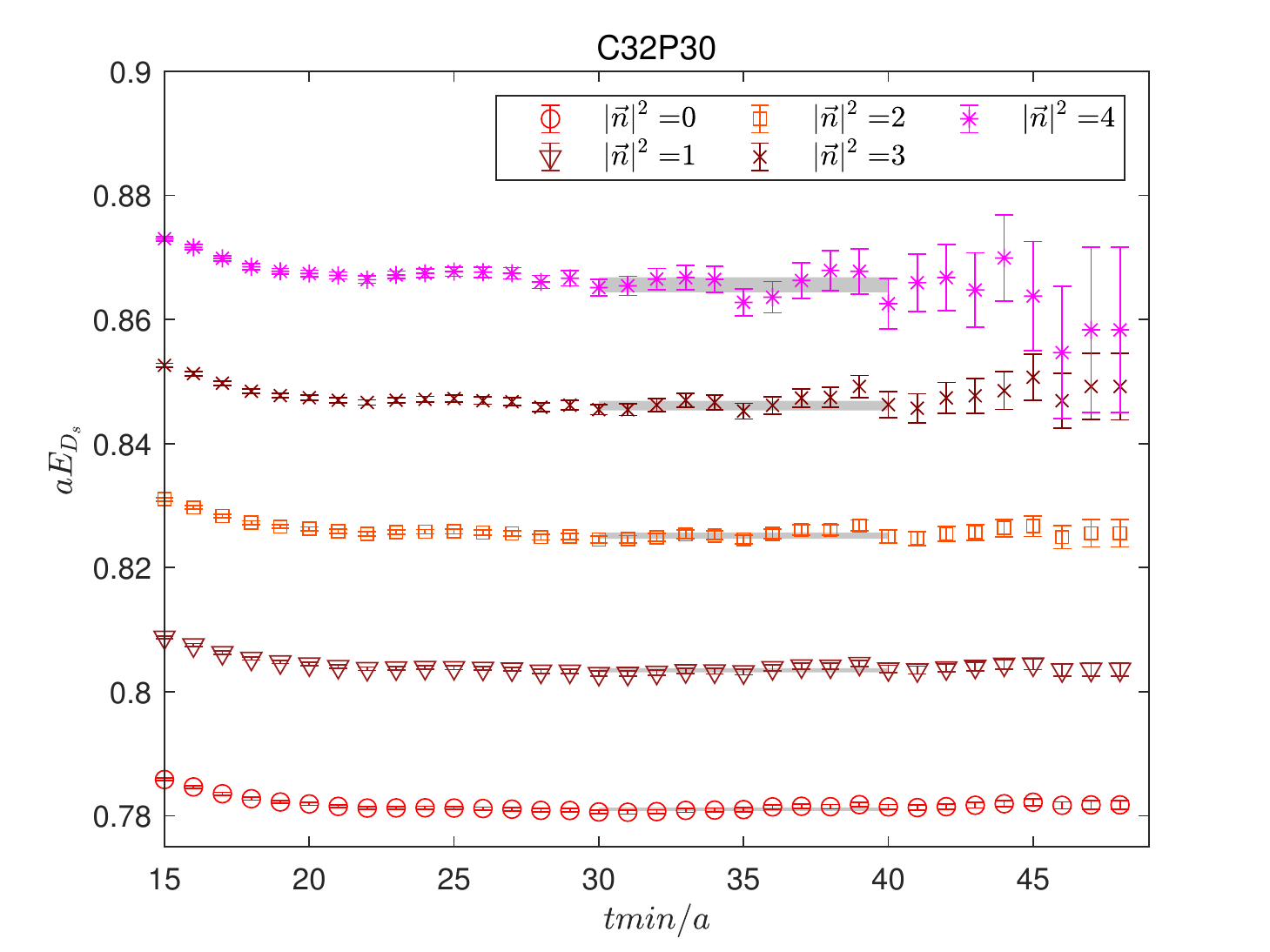}
}
\subfigure{
\centering
\includegraphics[width=5.5cm]{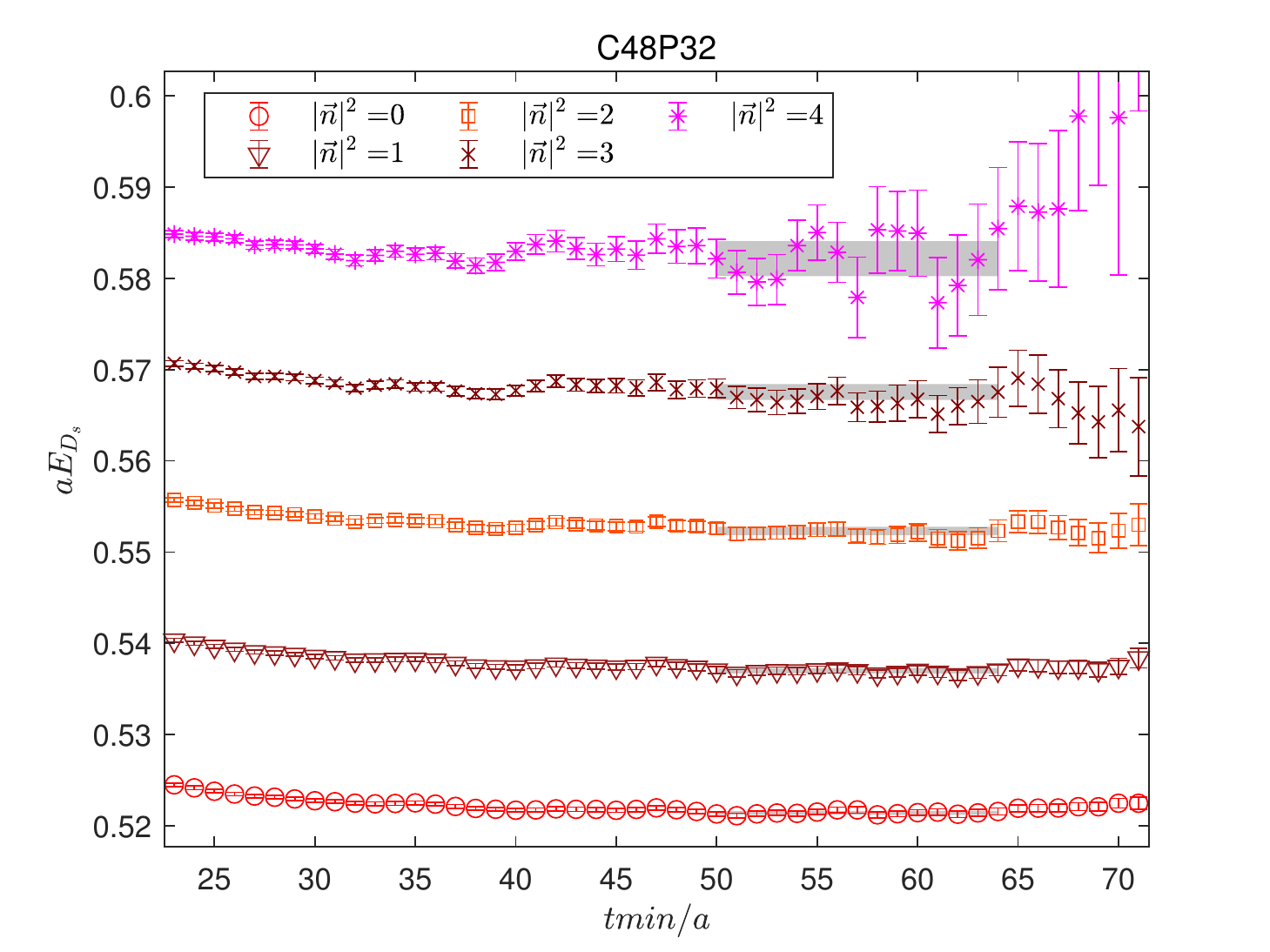}
}
\caption{\label{diag:mass} For the ensembles of C24P29, C32P30, and C48P32 from left to right, the mass spectra of $D_s^*$ and $D_s$ particles extracted from the two-point functions calculated using wall source propagators (top) and the energy levels of $D_s$ particle with different momentum $\vec{p}=2\pi\vec{n}/L,|\vec{n}|^2=0,1,2,3,4$ extracted from two-point functions calculated using point source propagators (bottom). The horizontal gray bands denote the fitting ranges.}
\end{figure*}

We employ three (2+1)-flavor Wilson-clover gauge ensembles generated by the CLQCD collaboration with lattice spacings $a\approx  0.1053,0.0775,0.0519$ fm, the parameters of which are shown in Table.~\ref{table:cfg}. For more details, we refer to Ref.~\cite{CLQCD:2023sdb}. A very fine lattice spacing with 0.03fm has been produced by MILC collaboration using the tadpole-improved symanzik gauge action~\cite{FermilabLattice:2018est}. Therefore, a similar setup with lattice spacing 0.052fm should be quite conservative to avoid the topology freezing effect. All the ensembles have similar volumes and pion masses in physical units and are expected to provide a fully well-controlled continuous extrapolation. The bare valence charm quark mass has not been presented in the original paper, so we determine its value by demanding the lattice result of $J/\psi$ mass to reproduce its physical value. This is due to the fact that the annihilation effect of $J/\psi$ particle is verified to be much smaller than the $\eta_c$ meson~\cite{Levkova:2010ft}. The latter is expected to cause a $1-4$ MeV mass shift.

\subsection{Mass spectrum}\label{sec:mass_spectrun}
\begin{table}[!h]
\begin{ruledtabular}
\begin{tabular}{cccc}
Ensemble & C24P29 & C32P30 &C48P32 \\
\hline
$m_{D_s^*}$(MeV) & 2086.5(1.2) & 2098.1(1.2) & 2117.1(2.5) \\
$m_{D_s}$(MeV) & 1998.2(0.5) & 1989.4(0.5) & 1983.8(0.6) \\
$Z_{D_s^*}$ & 0.1474(9) & 0.0858(5) & 0.0414(7) \\
$Z_{D_s}$ & 0.2175(5) & 0.1388(3) & 0.0722(2) \\
\end{tabular}
\end{ruledtabular}
\caption{Mass spectra $m_{D_s^*/D_s}$ and overlap function $Z_{D_s^*/D_s}$ for $D_s^*$ and $D_s$ particles, which are extracted from the two-point function calculated by the wall source propagators. $\delta m\equiv m_{D_s^*}-m_{D_s}$ is the hyperfine splitting.}
\label{tab:meson_mass}
\end{table}

The ground-state energies of the particle $D_s$ and $D_s^*$ are extracted from the two-point functions which are calculated by the wall source propagators. It is found in our study that the uncertainty is reduced by $30\%\sim 60\%$ by using a wall propagator compared to that using the point source propagator. For the determination of energy levels especially with nonzero momenta, we calculated them directly by the point source propagators. A single-state correlated fit with the formula Eq.~(\ref{eq:2pt}) is utilized and the numerical fitting results of the spectra are summarized in the Table~\ref{tab:meson_mass} and Table~\ref{tab:disper}. The effective levels of the particle $D_s$ and $D_s^*$ are both shown in Fig.~\ref{diag:mass} for all the ensembles and the horizontal gray bands therein denote the fitting center values and statistical errors estimated by the jackknife method. The ground state masses are shown by the upper panels in Fig.~\ref{diag:mass} and the energy levels with a series of momenta are illustrated in the lower panels.

\begin{figure}[htbp]
\centering
\includegraphics[width=8.5cm]{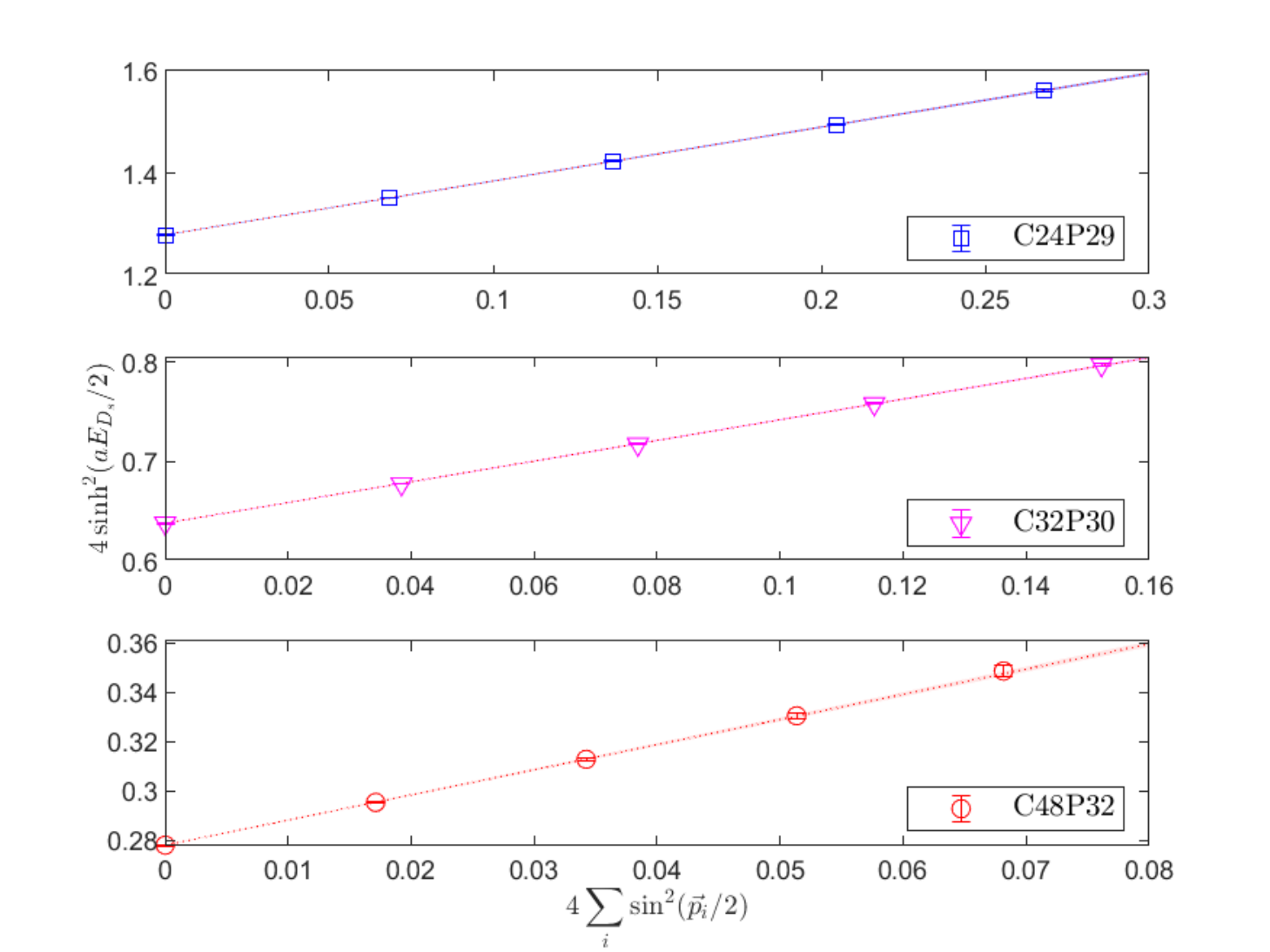}
\caption{\label{diag:disper} The dispersion relation of $D_s$ meson for the ensembles C24P29, C32P30, and C48P32, respectively. }
\end{figure}

We also check the dispersion relation of $D_s$ particle using the energy levels summarized in Table~\ref{tab:disper}. This verification is crucial since the energy of $D_s$ at nonzero momenta, namely, $E_{D_s}$, directly enters our calculation of the transition factor in Eq.~(\ref{eq:V_eff}). It is found that the discrete
dispersion relation
\be
4\sinh^2\frac{E_{D_s}}{2} =4\sinh^2\frac{m_{D_s}}{2}+\mathcal{Z}_{\textrm{latt}}^{D_s}\cdot 4\sum\limits_i \sin^2 \frac{\vec{p}_i}{2}
\ee
describes the energies and momenta well and a nice linear behavior between $4\sinh^2 (E_{D_s}/2)$ and $4\sum\limits_i \sin^2 (\vec{p}_i/2)$ is obtained as illustrated in Fig.~\ref{diag:disper}. The numerical values of the slope are well consistent with one, leading to a well-satisfying discrete dispersion relation in our simulations.

\begin{table}[!h]
\center
\begin{ruledtabular}
\begin{tabular}{cccccc}
Ensemble & C24P29 & C32P30 &C48P32 \\
\hline
$aE_{D_s}(|\vec{n}|^2=0)$ & 1.0657(4) & 0.7811(3) & 0.5214(2)\\
$aE_{D_s}(|\vec{n}|^2=1)$ & 1.0926(5) & 0.8035(3) & 0.5370(3) \\
$aE_{D_s}(|\vec{n}|^2=2)$ &1.1187(6) & 0.8251(5)& 0.5523(5) \\
$aE_{D_s}(|\vec{n}|^2=3)$ & 1.1441(9) & 0.8461(8) & 0.5672(9) \\
$aE_{D_s}(|\vec{n}|^2=4)$ & 1.1671(14) & 0.8656(12) & 0.5821(18) \\
$\mathcal{Z}_{\textrm{latt}}^{D_s}$ & 1.0305(89) & 1.0230(90) & 1.0182(130) \\
\end{tabular}
\end{ruledtabular}
\caption{Numerical results $E_{D_s}(\vec{p})$ with $\vec{p}=2\pi\vec{n}/L,|\vec{n}|^2=0,1,2,3,4$. The coefficient $\mathcal{Z}_{\textrm{latt}}^{D_s}$ is defined by the discrete dispersion relation. All the results are extracted from the two-point function calculated by the point source propagators.}
\label{tab:disper}
\end{table}

\subsection{$D_s^{*} \rightarrow D_s \gamma$}\label{sec:radiative}

\begin{figure}[htbp]
\centering
\subfigure{
\centering
\includegraphics[width=8.5cm]{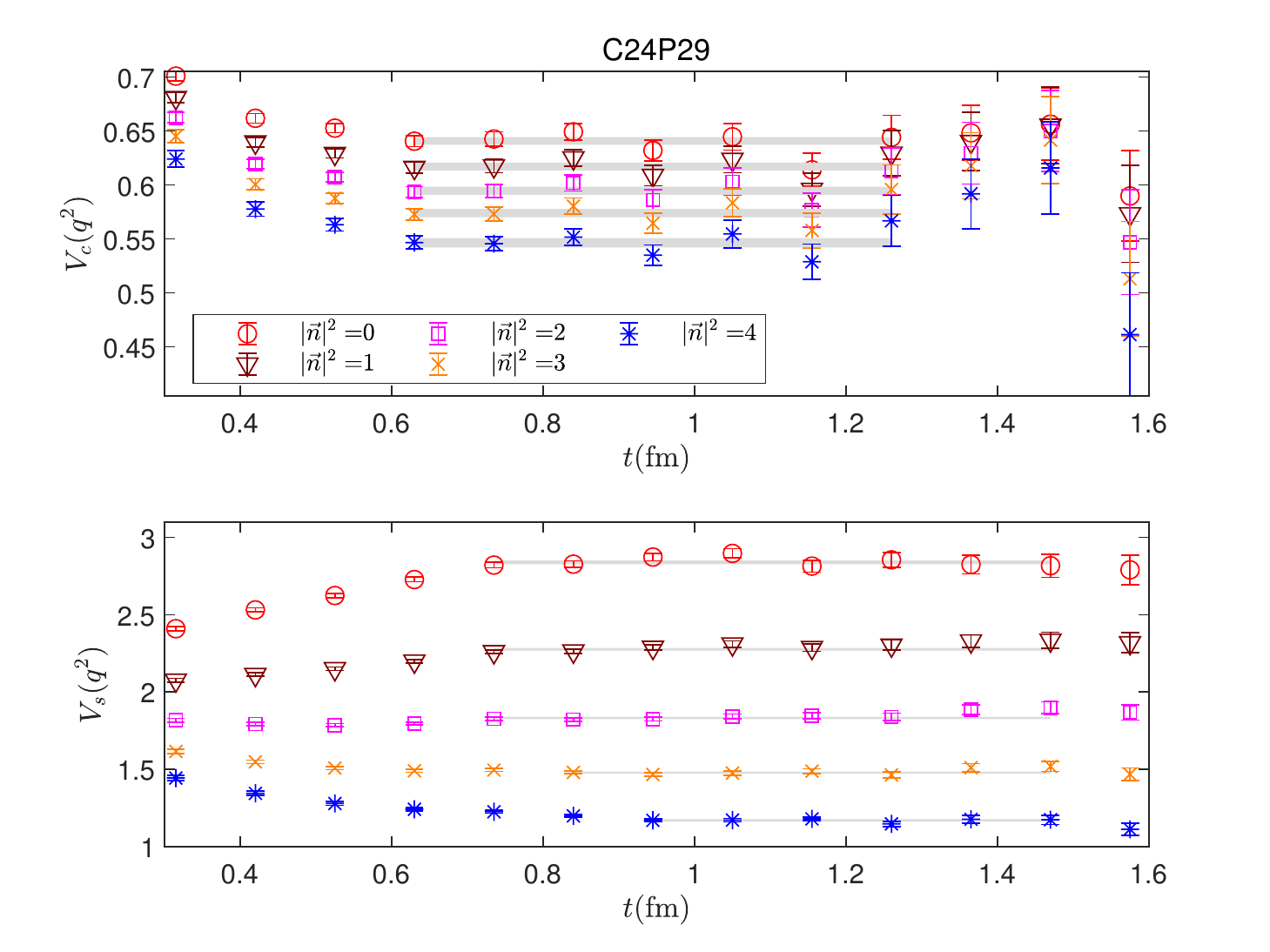}
}
\subfigure{
\centering
\includegraphics[width=8.5cm]{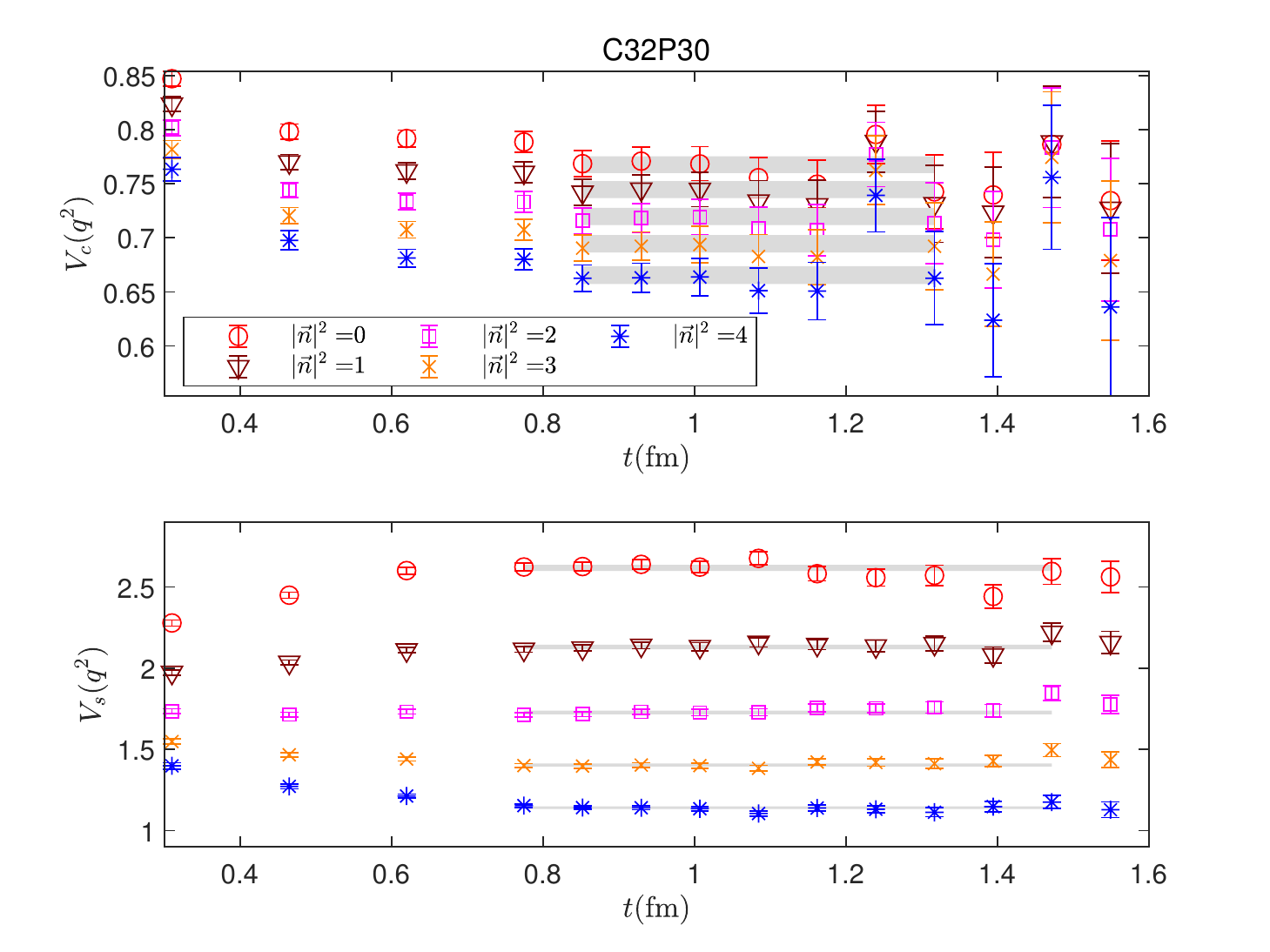}
}
\subfigure{
\centering
\includegraphics[width=8.5cm]{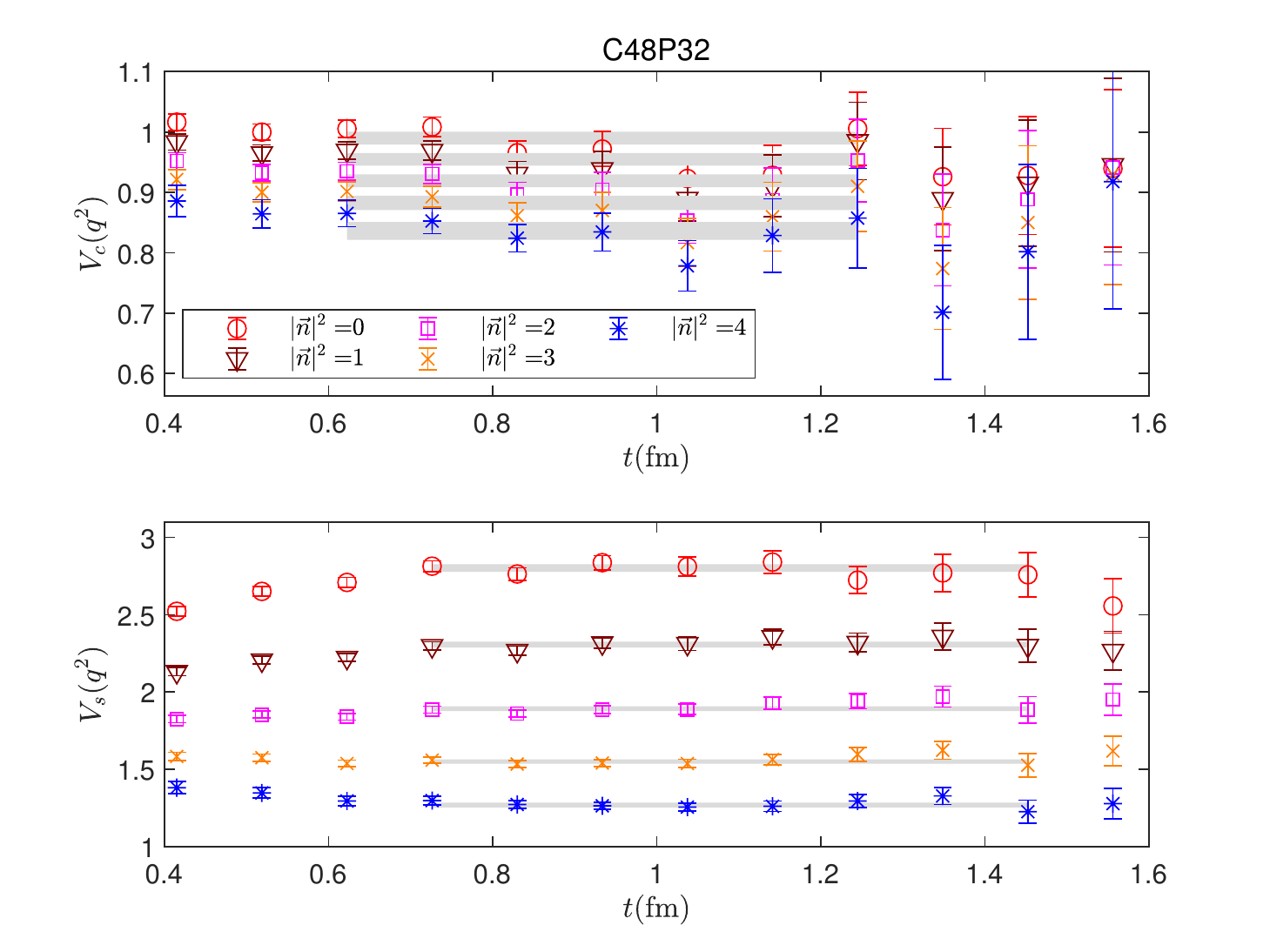}
}
\caption{\label{diag:Vcs} The transition factors $V_{c/s}(q^2)$ with different momentum $\vec{p}=2\pi\vec{n}/L,|\vec{n}|^2=0,1,2,3,4$. The horizontal gray bands denote the fitting results.}
\end{figure}

There is a total of two contributions for the effective transition factor $V_{\textrm{eff}}(q^2)$, one is that the photon is radiated from the charm quark, and the other is from a strange quark. This can also be read directly from the Wick contraction in Eq.~(\ref{eq:3pt_lat}). In the following, we divide the effective transition factor into two parts, $V_c(q^2)$ and $V_s(q^2)$. The former denotes the photon has escaped from the charm quark, and the latter from the strange quark. Specifically, it has

\be\label{eq:Veff}
V_{\textrm{eff}}(q^2)=\frac{1}{3}V_s(q^2)-\frac{2}{3}V_c(q^2)
\ee

Since these two parts come from different Wick contractions, each of them can be calculated separately. The lattice results of $V_{c}(q^2)$ and $V_{s}(q^2)$ as a function of the time separation $t$ are shown in Fig.~\ref{diag:Vcs}, together with a series of momenta $\vec{p}=2\pi\vec{n}/L,|\vec{n}|^2=0,1,2,3,4$.
Here we present all the results from the three ensembles, i.e. C24P29, C32P30, and C48P32 from top to bottom, respectively. It shows that $V_{c}(q^2)$ and $V_s(q^2)$ have obvious $t$ dependence in the small time region, indicating sizable
excited-state effects associated with the initial and final hadrons. With enough time intervals utilized in this work, we could observe obvious plateaus in a large enough time region. Therefore, excited-state effects are well controlled in our calculations. All results of $V_{c}(q^2)$ and $V_s(q^2)$ are obtained by a correlated fit of the lattice data to a constant at a suitable time region, which is denoted by the gray bands in the figure.
The fitting values are summarized in Table.~\ref{tab:Vcs_values}.

\begin{table}[!h]
\center
\begin{ruledtabular}
\begin{tabular}{cccccc}
Ensemble & C24P29 & C32P30 &C48P32 \\
\hline
$V_c(|\vec{n}|^2=0)$ & 0.6406(35) & 0.7675(78) & 0.9898(106) \\
$V_c(|\vec{n}|^2=1)$ & 0.6170(37) & 0.7445(78) & 0.9547(106) \\
$V_c(|\vec{n}|^2=2)$ & 0.5945(36) & 0.7197(79)& 0.9192(107)\\
$V_c(|\vec{n}|^2=3)$ & 0.5738(38) & 0.6945(80) & 0.8825(118) \\
$V_c(|\vec{n}|^2=4)$ & 0.5464(43) & 0.6655(82)& 0.8361(149) \\
\hline
$V_s(|\vec{n}|^2=0)$ & 2.8381(123) & 2.6166(183) & 2.8021(246) \\
$V_s(|\vec{n}|^2=1)$ & 2.2767(90) & 2.1297(142) &2.3074(190) \\
$V_s(|\vec{n}|^2=2)$ & 1.8327(72) & 1.7270(115) & 1.8928(156) \\
$V_s(|\vec{n}|^2=3)$ & 1.4789(69) & 1.4039(95) & 1.5519(144) \\
$V_s(|\vec{n}|^2=4)$ & 1.1724(69) & 1.1411(82) & 1.2703(154)\\
\hline
$V_{\textrm{eff}}(|\vec{n}|^2=0)$ & 0.5190(41) & 0.3605(58) & 0.2742(79) \\
$V_{\textrm{eff}}(|\vec{n}|^2=1)$ & 0.3476(32) & 0.2136(50) &0.1327(68) \\
$V_{\textrm{eff}}(|\vec{n}|^2=2)$ & 0.2146(28) & 0.0959(47) & 0.0182(64) \\
$V_{\textrm{eff}}(|\vec{n}|^2=3)$ & 0.1104(28) & 0.0050(47) & -0.0711(67) \\
$V_{\textrm{eff}}(|\vec{n}|^2=4)$ & 0.0265(31) & -0.0633(48) & -0.1340(86)\\
\end{tabular}
\end{ruledtabular}
\caption{Numerical results of $V_{c}(q^2)$, $V_{s}(q^2)$, and $V_{\textrm{eff}}(q^2)$ with $\vec{p}=2\pi\vec{n}/L,|\vec{n}|^2=0,1,2,3,4$.}
\label{tab:Vcs_values}
\end{table}

\begin{figure}[!h]
\centering
\subfigure{
\centering
\includegraphics[width=8.5cm]{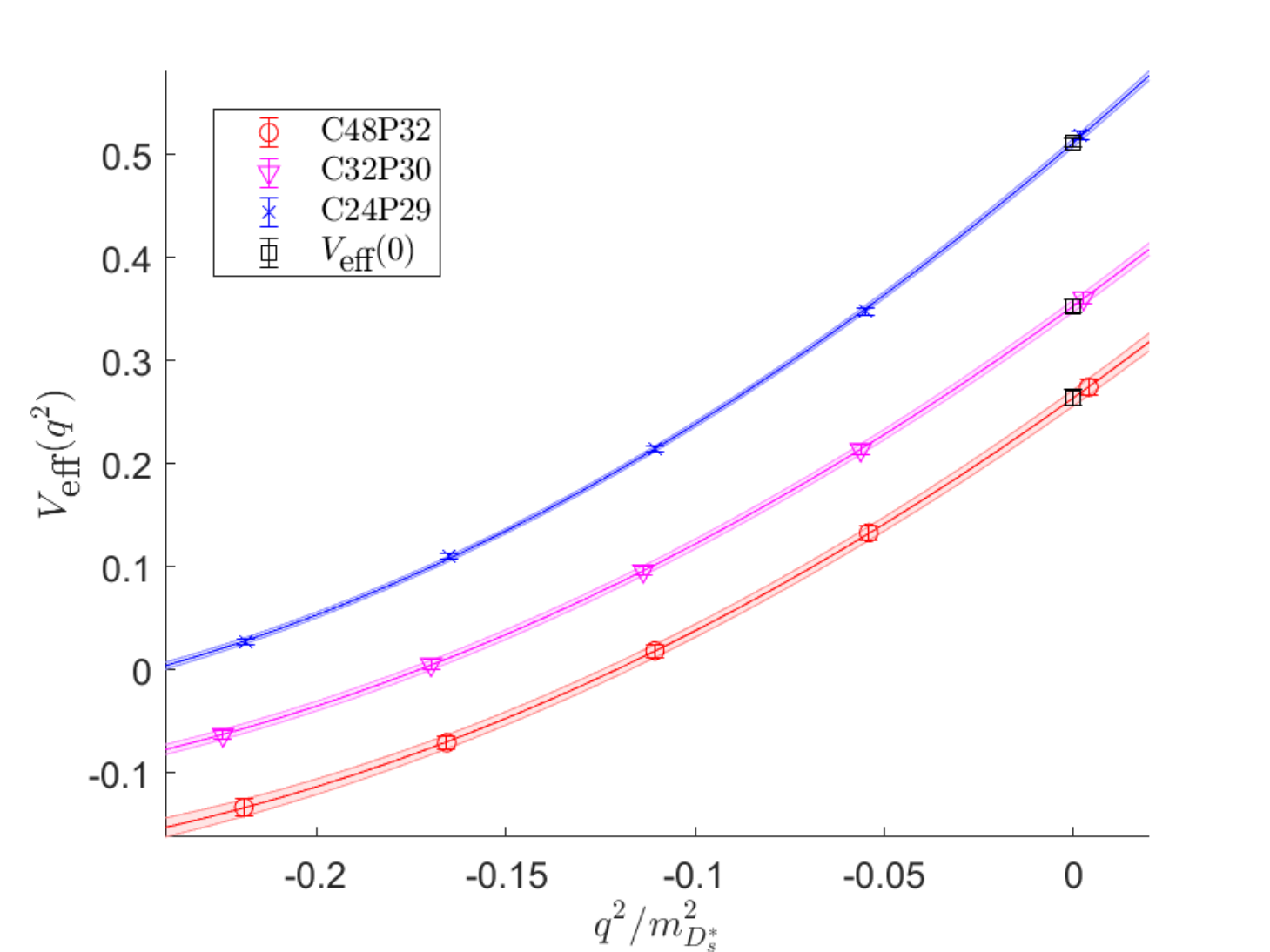}
}
\caption{\label{diag:Vcs_cont} The momentum extrapolation for transition factors $V_{\textrm{eff}}(q^2)$ with different momentum $\vec{p}=2\pi\vec{n}/L,|\vec{n}|^2=0,1,2,3,4$. The black squares denote the on-shell transition factor $V_{\textrm{eff}}(0)$.}
\end{figure}

Combining the transition factors $V_{c}(q^2)$ and $V_{s}(q^2)$, the effective transition factor $V_{\textrm{eff}}(q^2)$ can be calculated immediately, the numerical values of which are summarized in Table~\ref{tab:Vcs_values}. The on-shell factor $V_{\textrm{eff}}(0)$ is then extracted by a momentum extrapolation in $q^2\rightarrow 0$ with the five momentum modes taken into account. The numerical results of the coefficients $d_0,d_1$ and $d_2$ are summarized in Table~{\ref{tab:Vcs_d_values}}. In this work, it is found that the polynomial formula in Eq.~(\ref{eq:mom_extra}) can describe the lattice data very well, as shown in Fig.~\ref{diag:Vcs_cont}. It is seen that due to similar masses of the initial and final particles, i.e. $D_s^{*}$ and $D_s$, the on-shell transition factor is very close to that with zero momentum. Therefore, the statistical error of the on-shell transition factor is almost dominated by the error of the off-shell transition factor with zero momentum.

\begin{table}[!h]
\center
\begin{ruledtabular}
\begin{tabular}{ccccc}
Ensemble & C24P29 & C32P30 &C48P32 \\
\hline
$d_0$ & 0.512(4) & 0.353(6) & 0.264(8)\\
$d_1$ & 3.177(30)& 2.682(34) & 2.638(58)\\
$d_2$ & 4.419(103) & 3.693(88) & 3.746(242)\\
\end{tabular}
\end{ruledtabular}
\caption{Numerical results of $d_i$ with $i=0,1,2$ are extracted by the momentum extrapolation in Eq.~(\ref{eq:mom_extra}).}
\label{tab:Vcs_d_values}
\end{table}

\begin{figure}[!h]
\centering
\subfigure{
\centering
\includegraphics[width=8.5cm]{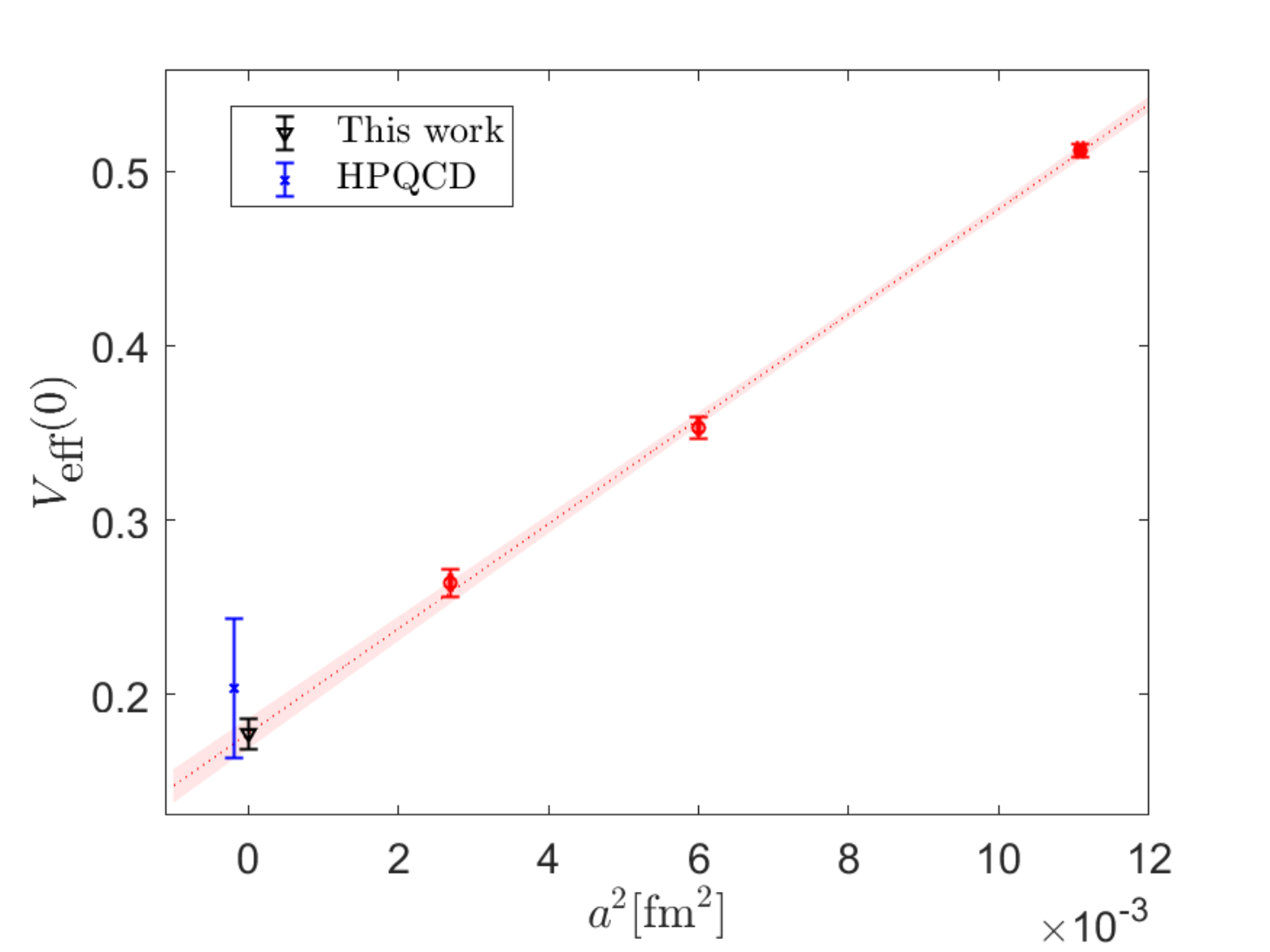}
}
\caption{\label{diag:decay_width} The lattice results of $V_{\textrm{eff}}(0)$ as a function of the lattice spacing. The errors of lattice spacing have been included in the continuous limit, which are presented by the horizontal error bars. The symbols of the blue cross and black triangle indicate the results given by HPQCD and this work, respectively.}
\end{figure}

The lattice results for the on-shell effective transition factor at different lattice spacings are shown in Fig.~\ref{diag:decay_width}, together with an extrapolation that is linear in $a^2$. This linear behavior is expected since the ensembles used in the work have adopted the tadpole-improved tree-level Symanzik gauge action and the tadpole-improved tree-level Clover fermion action. It is also seen that the fitting curves describe the lattice data well. After the continuous extrapolation, we obtain
\be
V_{\textrm{eff}}(0)=0.178(9)
\ee

Compared to the previous lattice calculation given by HPQCD~\cite{Donald:2013sra}, the charm quark in our simulation still contains a relatively large discretization error. However, with several improvements in our calculation, for example, the scalar function methodology, the especially finer lattice spacings, and many time separation utilized, we finally obtain $V_{\textrm{eff}}(0)$ with a significantly improved precision, where the statistical error is more than 4 times smaller than before. Using the physical transition factor $V_{\textrm{eff}}(0)=0.178(9)$ as input, and considering the physical masses of $D_s^*$ and $D_s$, i.e. $m_{D_s^*}=2112.2(4)$ MeV, $m_{D_s}=1968.35(7)$ MeV~\cite{10.1093/ptep/ptac097}, the decay width of the radiative decay
$D_s^* \rightarrow \gamma D_s$ appears to be\
\be\label{eq:decay_width_value}
\Gamma(D_s^*\rightarrow \gamma D_s) = 0.0549(54) ~\textrm{keV}
\ee
where the error only comes from the statistical error of the transition form factor $V_{\textrm{eff}}(0)$. For the first time, the accuracy of lattice calculation has reached percent level. It is seen our current result is also consistent with previous lattice calculations except that the statistical error is only a fifth of the previous ones. 

There are of course systematic errors that have not been 
seriously considered in this work. These include the effects from the neglected disconnected diagrams, the quenching of the charm quark, nonphysical light quark masses and finite volume effects. These effects could be studied in future systematic lattice studies using e.g. the gauge ensembles with physical pion mass, with charm sea quarks, and with more lattice spacings and volumes. In particular, the method proposed in this work has the potential to address the challenging contribution from the disconnected diagrams in the future.

With the input of the branching fraction $\operatorname{Br}(D_s^*\rightarrow D_s\gamma)=93.5(7)\%$, it immediately obtains the total decay width $\Gamma^{\textrm{total}}_{D_s^*}=0.0587(54)$ keV. Recently, the BESIII has reported the first experimental study of the purely leptonic decay $D_s^{+,*}\rightarrow e^+\nu_{e}$, and gives the branching fraction of this decay as $(2.1^{+1.2}_{-0.9_{\textrm{stat.}}}\pm 0.2_{\textrm{syst.}})\times 10^{-5}$~\cite{BESIII:2023zjq}. Combining this branching fraction with our lattice calculation on the total decay width, we obtain $f_{D_s^*}|V_{cs}|=(190.5^{+55.1}_{-41.7_{\textrm{stat.}}}\pm 12.6_{\textrm{syst.}})$ MeV, where the stat. is only the statistical error from the experiment, and syst. results from the experimental systematic uncertainty and lattice statistical error. If take the previous lattice QCD prediction of the total decay width $0.070(28)$ keV as input, it leads to a systematic error $42.7_{\textrm{syst.}}$ for the quantity $f_{D_s^*}|V_{cs}|$. At present, the updated systematic uncertainty is mainly from the uncertainties in the measured $\operatorname{Br}(D_s^{+,*}\rightarrow e^+\nu_{e})(9.5\%)$ and the LQCD updated $\Gamma^{\textrm{total}}_{D_s^*}(9.2\%)$.

\subsection{$D_s^{*} \rightarrow D_s e^{+}e^{-}$}\label{sec:dalitz}

\begin{figure}[!h]
\centering
\subfigure{
\centering
\includegraphics[width=8.5cm]{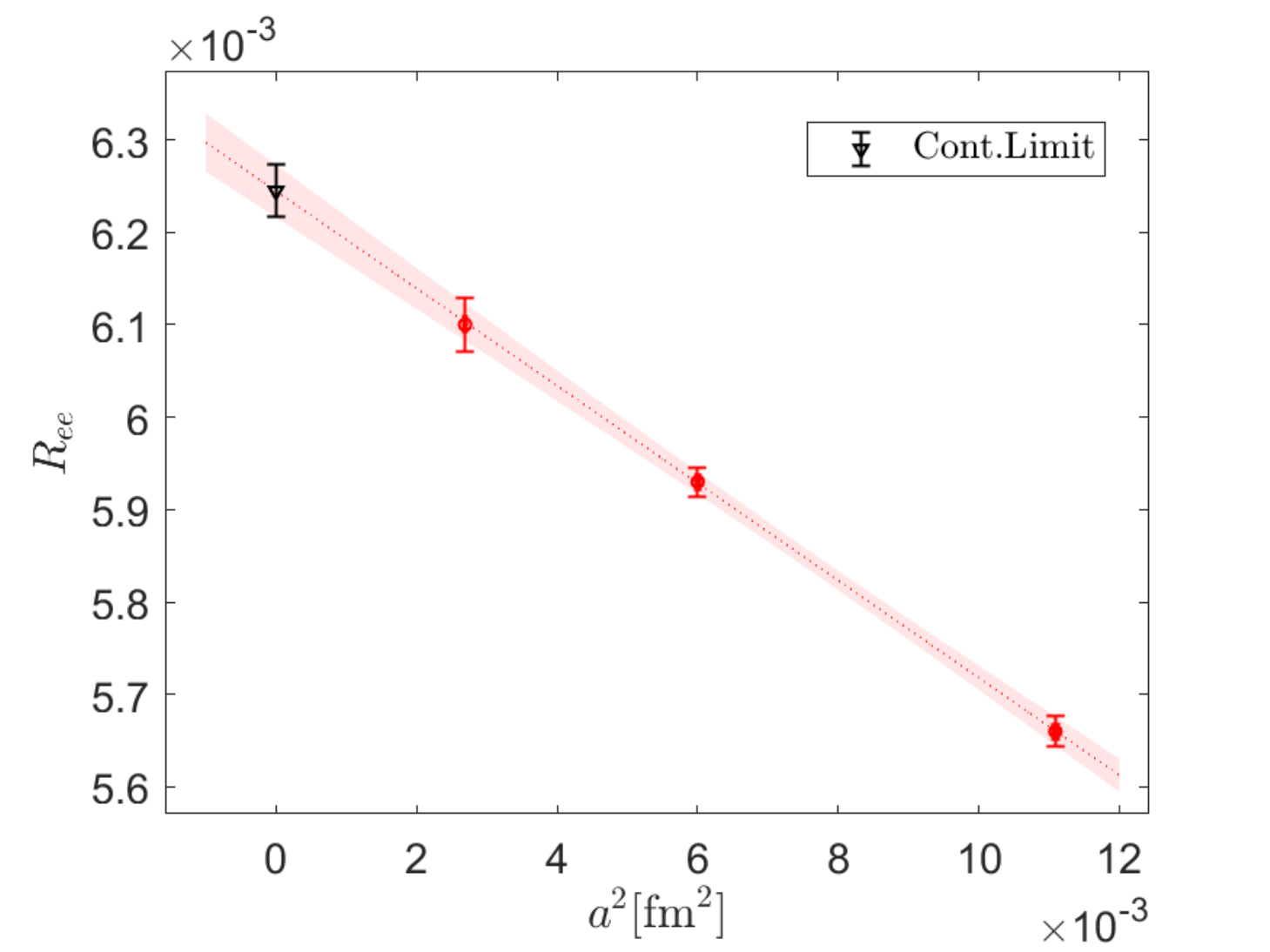}
}
\caption{\label{diag:Ree} The lattice results of $R_{ee}$ as a function of the lattice spacing. The black triangle denotes the result in the continuous limit $a^2\rightarrow 0$.}
\end{figure}

\begin{table}[!h]
\center
\begin{ruledtabular}
\begin{tabular}{cccccc}
Ensemble & C24P29 & C32P30 &C48P32 & Cont.Limit \\
\hline
$R_{ee}\times 10^3$ & 5.66(2) & 5.93(2) & 6.10(3) & 6.24(3)\\
\end{tabular}
\end{ruledtabular}
\caption{Numerical value of $R_{ee}$ from three gauge ensembles, together with the physical result in a continuous limit.}
\label{tab:Ree}
\end{table}

For the Dalitz decay of $D_s^*$, a virtual photon is internally converted to a leptonic pair $l^+l^-$. Since the $\mu$ mass is larger than the mass splitting of $D_s^*$ and $D_s$, the only possible decay mode is the $e^+e^-$ pair. Taking into account the transition factor $V_{\textrm{eff}}(q^2)$ already obtained in the calculation of $D_s^*\rightarrow D_s\gamma$, the ratio $R_{ee}$ defined in Eq.~(\ref{eq:Ree}) is calculated straightforwardly. The results of the ratio for all ensembles are shown in Fig.~\ref{diag:Ree} and the numerical values are presented in Table~\ref{tab:Ree}. A linear behavior in $a^2$ can well describe the lattice data as expected. Finally, we have
\be
R_{ee}=\frac{\Gamma(D_s^*\rw D_se^+e^-)}{\Gamma(D_s^*\rw D_s\gamma)}=0.624(3)\%
\ee
which is consistent with the PDG value $0.67(16)\%$. The errors of the subtracting transition factors $V_{\textrm{eff}}(q^2)$ and $V_{\textrm{eff}}(0)$ almost completely cancel, eventually leading to a per mill level calculation.

As far as we know, three main decay channels of $D_s^*$ are listed by PDG, and they are $D_s^*\rightarrow D_s\gamma$, $D_s^*\rightarrow D_s \pi^0$, and $D_s^*\rightarrow D_se^+e^-$. Their respective branching fractions are determined from only two relative ratios, i.e. $R_{ee}=\Gamma(D_s^*\rightarrow D_se^+e^-)/\Gamma(D_s^*\rightarrow D_s\gamma)$ and $R_{Ds\pi^0}=\Gamma(D_s^*\rightarrow D_s\pi^0)/\Gamma(D_s^*\rightarrow D_s\gamma)$. By assuming the sum of the three branching fractions is equal to 1, one has $\operatorname{Br}(D_s^*\rightarrow D_s\gamma)=1/(1+R_{ee}+R_{D_s\pi})$, $\operatorname{Br}(D_s^*\rightarrow D_s \pi^0)=R_{D_s\pi^0}/(1+R_{ee}+R_{D_s\pi})$, and $\operatorname{Br}(D_s^*\rightarrow D_se^+e^-)=R_{ee}/(1+R_{ee}+R_{D_s\pi})$. It is seen that the quantity $R_{ee}$ is directly measurable. Therefore, such a physical quantity especially with ultra-high precision serves as an excellent ground for testing the standard model. More accurate experimental measurements on $R_{ee}$ are welcome in the future.

\section{Conclusion}\label{sec:conclude}
In this work, we present a lattice QCD calculation on the radiative decay of $D_s^*$ particle. The transition process $D_s^*\rightarrow D_s\gamma$ and Dalitz decay $D_s^*\rightarrow D_s e^+e^-$ are studied respectively. Using the 2+1 Wilson Clover gauge ensembles under three different lattice spacings, we finally obtain $\Gamma(D_s^*\rightarrow D_s \gamma)=0.0549(54)$ keV, with an statistical error significantly reduced compared to previous lattice calculation. The Dalitz decay of $D_s^*$ is also studied for the first time, and the ratio is obtained as $R_{ee}=0.624(3)\%$, with a precision better than one percent. 

To reach a percent level calculation, further improvements in several aspects are adopted. First, we utilize a scalar function method to calculate the effective transition factor of $D_s^*\rightarrow D_s\gamma$, where the zero transfer momentum can be projected directly without any ambiguity. As the mass splitting of $D_s^*$ and $D_s$ is relatively small, the on-shell transition factor is very close to that with zero momentum. Therefore, the precision of the on-shell transition factor is almost dominated by the off-shell factor with zero momentum after a momentum extrapolation. Second, a large number of time separations $t$ have been utilized in our calculation, the excited-state contamination caused by the initial and final states are therefore removed and the transition factor is obtained by a correlated fit to a constant at large $t$. Third, we have used three ensembles with different lattice spacings to perform a continuous limit, especially including a very fine spacing with only 0.052 fm. Taking into
account of the above-mentioned improvements, we managed to obtain a result for the decay width in Eq.~(\ref{eq:decay_width_value}) with an statistical error of about 9.8\%.

A precise determination of $D_s^*$ total decay width plays a vital role in extracting the CKM matrix element $V_{cs}$. Combining with a recent experimental measurement on $D_s^{*,+}\rightarrow e^+\nu_e$, the quantity $f_{D_s^*}|V_{cs}|$ is estimated and found to be $(190.5^{+55.1}_{-41.7_{\textrm{stat.}}}\pm 12.6_{\textrm{syst.}})$ MeV. The systematic error here is significantly reduced compared to 42.7 which is extracted using the previous lattice result $\Gamma_{D_s^*}^{\textrm{total}}=0.070(28)$ keV as input. A further improvement on the measurement of $D_s^*$ purely leptonic decay, both statistically and systematically, is expected for the next generation of the 
supercollider, such as Super Tau Charm Facility~\cite{Achasov:2023gey}.
\begin{acknowledgments}
We gratefully acknowledge the helpful discussions with CLQCD members. Y.M. thanks Liuming Liu, Wei Sun, and Yi-Bo Yang for providing the helps on the chroma software~\cite{Edwards:2004sx}. Y.M. and C.L. are supported by NSFC of China under Grant No.12293060, No.12293063, No. 12305094, No.11935017, and No.12070131001. Z. L. is supported by NSFC of China under Grant No. 12075253 and No. 12192264.
The calculation is supported by SongShan supercomputer at the National Supercomputing Center in Zhengzhou. T. S. is supported by Joint Fund of Research utilizing Large-Scale Scientific Facility of the NSFC and CAS under Contract
No. U2032114.
\end{acknowledgments}

\bibliography{ref}

\bibliographystyle{h-physrev}

\end{document}